\documentclass[fleqn,12pt,twoside]{article}
\usepackage{espcrc1}
\usepackage{graphics}

\usepackage{graphicx}
\usepackage[figuresright]{rotating}

\newcommand{\AmS}{{\protect\the\textfont2
  A\kern-.1667em\lower.5ex\hbox{M}\kern-.125emS}}

\title{Measurement of Source Chaoticity for Particle Emission in Au+Au
Collisions at $\sqrt{s_{NN}}$ = 130 GeV using 3-Particle HBT Correlations}

\author{R. Willson\address[]{The Ohio State University, 
        174 W. 18th Ave., Columbus, Ohio 43210, USA} for the STAR
	Collaboration\thanks{For the full author list and acknowledgements, see Appendix
        ``Collaborations'' of this volume.}}
       
\begin{document}

% typeset front matter
\maketitle

\begin{abstract}
Data from the first physics run at the Relativistic Heavy-Ion Collider
at Brookhaven National Laboratory from the STAR experiment have been
analyzed using three-pion correlations to study whether pions are
emitted independently at freezeout. We have made a high-statistics
measurement of the three-pion correlation function and calculated
the normalized three-particle correlator to obtain a quantitative
measurement of the degree of chaoticity in the freeze-out environment. 
\end{abstract}

\section{Introduction}

Two-pion Hanbury Brown and Twiss (HBT) interferometry in principle
provides a means of extracting the space-time evolution of the pion
source at freeze-out produced in relativistic heavy-ion collisions
\cite{Gyulassy,Report}. An underlying assumption of this method is
that pions are produced from a completely chaotic source, i.e. a source
in which the hadronized pions are created with random quantum particle
production phases. Although two-pion HBT provides some estimation of
this chaoticity, a better method is available using 
three-particle correlations. Normalizing the three-pion correlation
function appropriately by the two-pion correlator, the effects from
particle misidentification and decay contributions can be made to
drop out \cite{HZ}, thereby isolating possible coherence effects
in the particle emission process. The resulting three-pion correlator
\( r_{3} \) provides the means of extracting the degree of source
chaoticity by examining its value at zero relative momentum.

\section{Derivation of Chaoticity from Normalized Three-pion Correlator}

The measured observable in this analysis is the normalized three-pion correlator \cite{HZ}:

\begin{equation}
\label{CosPhi}
r_{3}\left( Q_{3}\right) =\frac{\left( C_{3}\left( Q_{3}\right) -1\right) -\left( C_{2}\left( Q_{12}\right) -1\right) -\left( C_{2}\left( Q_{23}\right) -1\right) -\left( C_{2}\left( Q_{31}\right) -1\right) }{\sqrt{\left( C_{2}\left( Q_{12}\right) -1\right) \left( C_{2}\left( Q_{23}\right) -1\right) \left( C_{2}\left( Q_{31}\right) -1\right) }}
\end{equation}

Here \( Q_{3}=\sqrt{Q_{12}^{2}+Q_{23}^{2}+Q_{31}^{2}} \)
and \( Q_{ij}=\sqrt{-(p_{i}{-}p_{j})^{2}} \) are the standard invariant
relative momenta which can be computed for each
pion triplet from the three measured momenta \( ({\textbf {p}}_{1},{\textbf {p}}_{2},{\textbf {p}}_{3}) \).
\( C_{2}\left( p_{1},p_{2}\right) =\frac{P_{2}\left( p_{1},p_{2}\right) }{P_{1}\left( p_{1}\right) P_{1}\left( p_{2}\right) } \)
and \( C_{3}\left( p_{1},p_{2},p_{3}\right) =\frac{P_{3}\left( p_{1},p_{2},p_{3}\right) }{P_{1}\left( p_{1}\right) P_{1}\left( p_{2}\right) P_{1}\left( p_{3}\right) } \),
where \( P \) represents the momentum probability distribution.

For fully chaotic sources \( r_{3} \) approaches \( 2 \) as all
relative momenta (and thus \( Q_{3} \)) go to zero. If the source
is partially coherent, a relationship can be established between the
limiting value of the three-pion correlator at \( Q_{3}=0 \) and
the chaotic fraction \( \varepsilon  \) (\( 0{\, \leq \, }\varepsilon {\, \leq \, }1 \))
of the single-particle spectrum \cite{HZ}: \begin{equation}
\label{Chaotic Fraction}
\frac{1}{2}\, r_{3}\left( Q_{3}{=}0\right) =\sqrt{\varepsilon }\, \frac{3-2\varepsilon }{(2{-}\varepsilon )^{3/2}}.
\end{equation}
Chaotic fraction $\varepsilon$ gives an upper limit on the value of the two-pion $\lambda$ parameter, which is
sensitive to the number of coherent pairs in a sample.

The three-boson correlation
function \( C_{3}\left( Q_{3}\right)  \) is calculated from the data
by taking the ratio \( \frac{A\left( Q_{3}\right) }{B\left( Q_{3}\right) } \)
and normalizing it to unity at large \( Q_{3} \). Here \( A\left( Q_{3}\right) =\frac{dN}{dQ_{3}} \)
is the three-pion distribution as a function of the invariant three-pion
relative momentum, integrated over the total momentum of the pion
triplet as well as all other relative momentum components. It is obtained
by taking three pions from a single event, calculating \( Q_{3} \),
and binning the results in a histogram. \( B\left( Q_{3}\right)  \)
is computed by taking a single pion from three separate events.

\section{Experimental Results and Discussion}

Data for the present results are from 1M events taken during the Year-1
physics run at STAR using the Time Projection Chamber (TPC) \cite{NIM}
as the primary tracking detector. Two multiplicity classes
were created by taking the 12\% most central for the high multiplicity
set, the next 20\% most central for the low multiplicity set. 
For both multiplicity bins, tracks were
constrained to have \( p_{T} \) in the range \( 0.125<p_{T}<0.5 \)\,GeV/\( c \),
and pseudorapidity \( \left| \eta \right| <1.0 \). In the range \( 0<Q_{3}<120 \)\,MeV/\( c \),
approximately 150 million triplets were included in both the negative
and positive pion studies.

The \( C_{2} \) correlation function was corrected for Coulomb repulsion
with a finite Gaussian source approximation \cite{Coulomb}. The \( C_{3} \),
correction factor is the product of three two-pion correction terms, obtained from the three pairs from the triplet.

In calculating \( r_{3} \),
the actual binned values of the correlation function for the various
values of \( Q_{3} \) are used instead of a fit \cite{TJH}. \( Q_{3} \), \( Q_{12} \),
\( Q_{23} \) and \( Q_{31} \) are calculated from triplets of particles from the dataset and
the three pairs that can be formed from the triplet. Eq. (\ref{CosPhi}) is then evaluated (as a function of \(Q_{3}\)) using the binned
two- and three-pion correlation functions, and averaged over the number of triplets in each \(Q_{3}\) bins.

\begin{figure}[htb]
\begin{minipage}[t]{80mm}
{\centering \resizebox*{1.0\columnwidth}{!}{\includegraphics{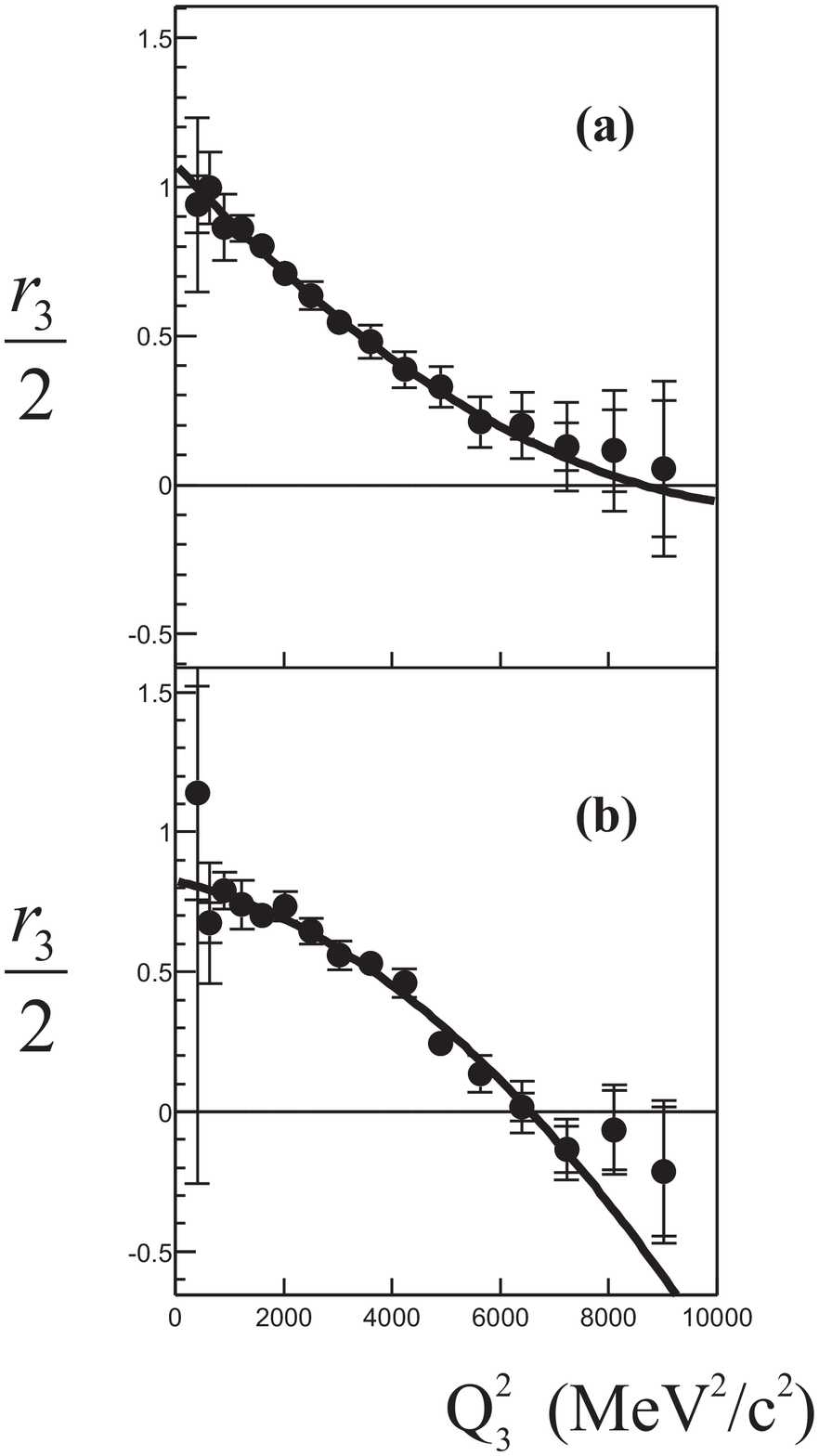}} \par}
\caption{\protect \protect\( r_{3}\protect \) calculation for (a) central and (b) mid-central
\protect\( \pi ^{-}\protect \) events.
The fits shown use Eq. \ref{Factorization} to determine the intercept.
Statistical and statistical+systematic errors are shown.}
{\centering \label{FIG:CosPhiNeg}\par}
\end{minipage}
\hspace{\fill}
\begin{minipage}[t]{75mm}
{\centering \resizebox*{1.06\columnwidth}{!}{\includegraphics{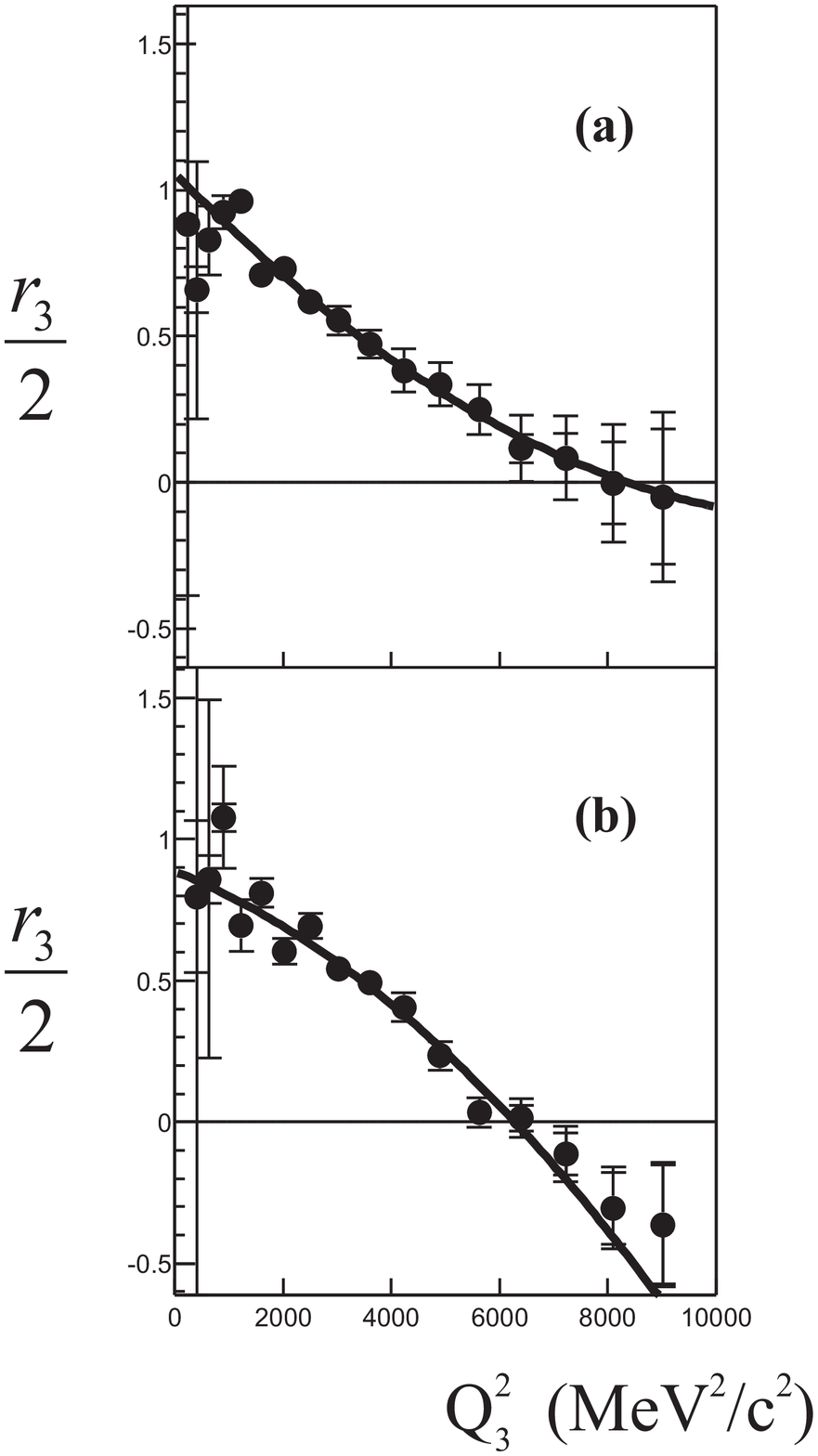}} \par}
\caption{\protect \protect\( r_{3}\protect \) calculation for (a) central and 
mid-central \protect\( \pi ^{+}\protect \) events.
The fits shown use Eq. \ref{Factorization} to determine the intercept.
Statistical and statistical+systematic errors are shown.}
{\centering \label{FIG:CosPhiPos}\par}
\end{minipage}
\end{figure}

The results for the three multiplicity bins are shown in Figures \ref{FIG:CosPhiNeg}
(\( \pi ^{-} \)) and \ref{FIG:CosPhiPos} (\( \pi ^{+} \)), plotted
as functions of \( Q_{3}^{2} \), and fitted to a function of the form:
\begin{equation}
\label{Factorization}
r_{3}\left( Q_{3}\right) =r_{3}(0)-C_{1}\, Q^{2}_{3}-C_{2}\, Q^{4}_{3}.
\end{equation}
This functional form is suggested by the theoretical
analysis in \cite{HZ} which shows that the leading relative momentum
dependencies in the numerator and denominator of Eq. \ref{CosPhi}
are in even powers of $Q_{3}$ \cite{fn1}.  The fit range used is \( 0<Q_{3}<120 \)\,MeV/\( c \).

\begin{figure}[htb]
\begin{minipage}[t]{80mm}
{\centering \resizebox*{0.92\columnwidth}{!}{\includegraphics{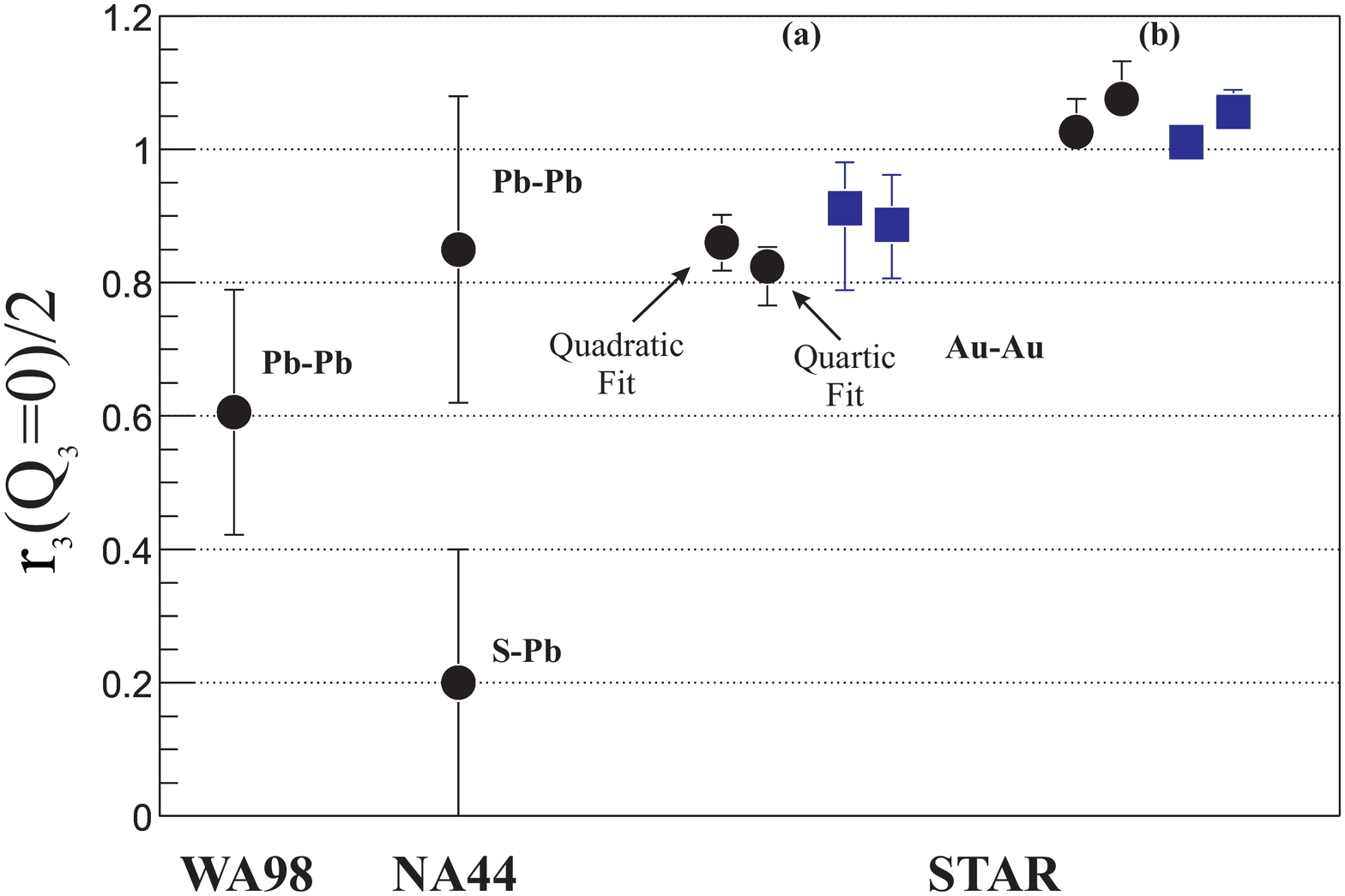}} \par}
\caption{\protect Asymptotic value of \protect\( r_{3}\protect \) from
STAR and two other experiments \cite{NA44,WA98}. For STAR, central
(a) and mid-central results are shown for \protect\( \pi ^{-}\protect \)
(circular markers) and \protect\( \pi ^{+}\protect \) (square markers)
data. For each STAR dataset, separate markers denote a fit using Eq. \ref{Factorization} (Quartic) and 
a fit to a quadratic equation (Quadratic). }
{\centering \label{FIG:SummaryCosPhi}\par}
\end{minipage}
\hspace{\fill}
\begin{minipage}[t]{75mm}
{\centering \resizebox*{1.0\columnwidth}{!}{\includegraphics{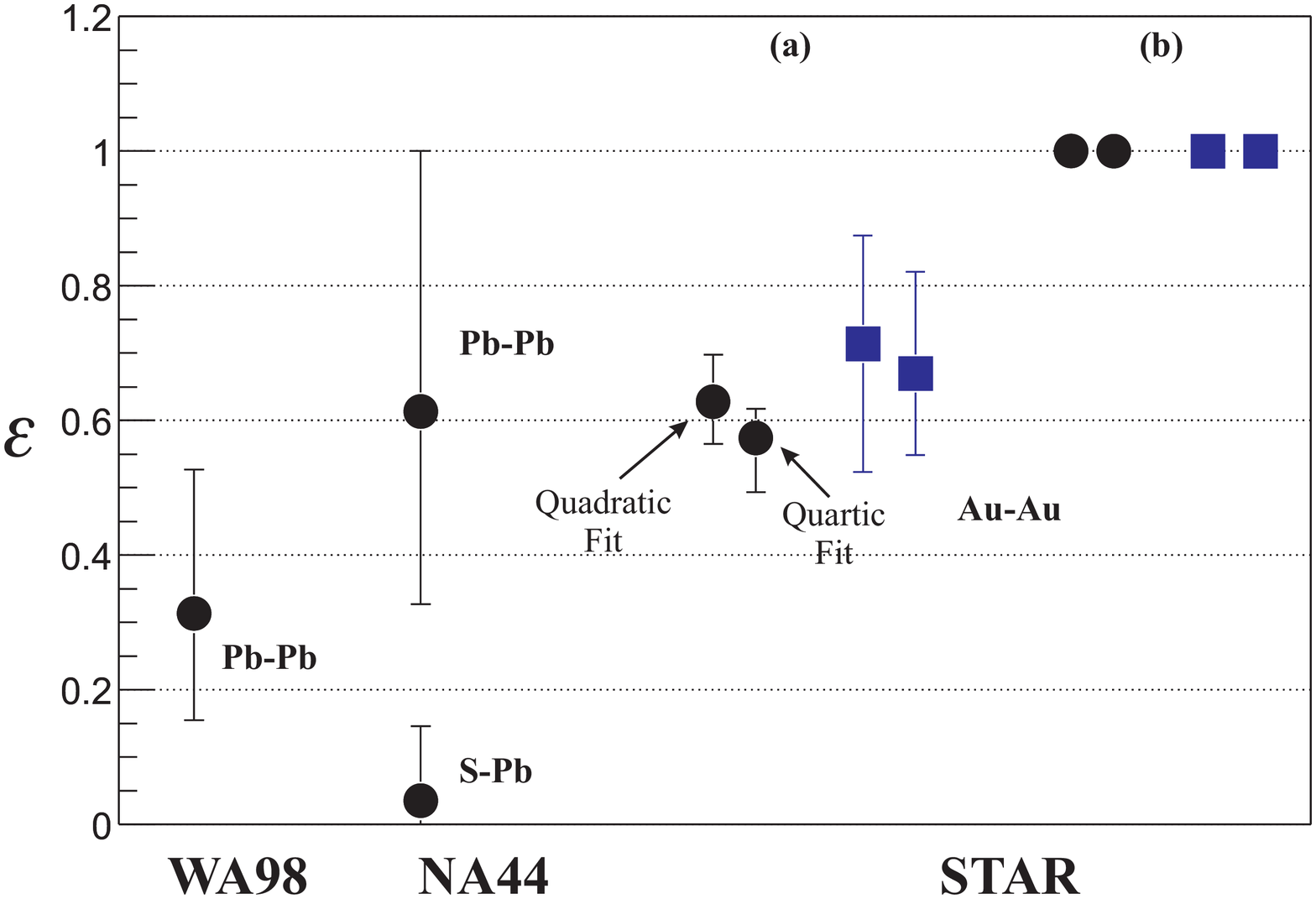}} \par}
\caption{\protect Chaotic fraction, calculated from Eq. \ref{Chaotic Fraction},
and plotted for the same experiments as in Figure \ref{FIG:SummaryCosPhi}.
The data markers are described in Figure \ref{FIG:SummaryCosPhi}.}
{\centering \label{FIG:SummaryCF}\par}
\end{minipage}
\end{figure}

The resulting intercepts \( r_{3}(0) \) are shown in Fig.~\ref{FIG:SummaryCosPhi},
along with the results of WA98 and NA44. NA44 reported a result close
to unity for Pb-Pb interactions, but a much lower result for S-Pb
\cite{NA44}, both with no clear \( Q_{3} \) dependence. WA98 also
reported a result close to unity at \( Q_{3}{\, =\, 0} \) for Pb+Pb,
and the \( Q_{3} \)-dependence in their result is similar to what
we see in central collisions \cite{WA98}.

Figure \ref{FIG:SummaryCF} shows the calculation of \( \varepsilon  \)
for STAR's measurements, and for those from WA98 and NA44. The plot
shows a systematic trend in the STAR results going
from the peripheral bin to the central bin, with the central results showing a fully
chaotic source. By calculating the number of coherent pair in the dataset from the chaotic fraction,
a maximum value of the two-pion $\lambda$ factor has been determined which is consistent with the 
value of $\lambda$ found in the two-pion analysis \cite{STARHBT}. 

\section{Conclusion}

In summary, we have presented three-pion HBT results for \( \sqrt{s_{NN}}{\, =\, }130 \)\,GeV
data at STAR, and have shown that for the two multiplicity classes the
STAR data indicate a large degree of chaoticity in the source at freeze-out.
High statistics from STAR have allowed a normalized three-pion correlator
calculation that extends to \( 120 \) MeV/\( c \) in \( Q_{3} \), and when used in
conjunction with the formalism of partially coherent sources obtained
from \cite{HZ}, quantitative limits on the fraction of chaoticity
are obtained which are in agreement with two-pion $\lambda$ parameter measured at STAR.
STAR's measured values provide increased confidence in the validity
of standard HBT analyses based on the assumption of a chaotic source.

\end{document}